\documentclass[useAMS,12pt]{article}

\usepackage{graphicx,epsfig,epsf,scicite}

\usepackage{times}

\def\hess{H.E.S.S.}
\def\ls{LS~5039}
\def\com#1{$^\dagger$}
\def\mathieucom#1{$^{\dagger\dagger}$}
\def\deg{$^{\rm o}$}

\newenvironment{sciabstract}{%
\begin{quote} \bf}
{\end{quote}}

\newcounter{lastnote}
\newenvironment{scilastnote}{%
\setcounter{lastnote}{\value{enumiv}}%
\addtocounter{lastnote}{+1}%
\begin{list}%
{\arabic{lastnote}.}
{\setlength{\leftmargin}{.22in}}
{\setlength{\labelsep}{.5em}}}
{\end{list}}

\title{Discovery of very high energy $\gamma$-rays associated with an X-ray binary}

\author{F.~Aharonian $^{1}$, 
  A.G.~Akhperjanian $^{2}$, 
  K.-M.~Aye $^{7}$,
  A.R.~Bazer-Bachi $^{3}$,
  M.~Beilicke $^{4}$,\\
  W.~Benbow $^{1}$,
  D.~Berge $^{1}$,
  P.~Berghaus $^{9}$,
  K.~Bernl\"ohr $^{1,5}$,
  C.~Boisson $^{6}$
% ...
\\
  \\
  \normalsize{$^\ast$To whom correspondence should be addressed; E-mail: denauroi@in2p3.fr, dubus@in2p3.fr}\\
}

\date{}

\begin{document} 




\maketitle

{\footnotesize
\begin{enumerate}
\item {Max-Planck-Institut f\"ur Kernphysik, P.O. Box 103980, D 69029
Heidelberg, Germany}
\item{Yerevan Physics Institute, 2 Alikhanian Brothers St., 375036 Yerevan,
Armenia}
\item{Centre d'Etude Spatiale des Rayonnements, CNRS/UPS, 9 av. du Colonel
Roche, BP 4346, F-31029 Toulouse Cedex 4, France}
\item{Universit\"at Hamburg, Institut f\"ur Experimentalphysik, Luruper Chaussee
149, D 22761 Hamburg, Germany}
\item{Institut f\"ur Physik, Humboldt-Universit\"at zu Berlin, Newtonstr. 15,
D 12489 Berlin, Germany}
\item{LUTH, UMR 8102 du CNRS, Observatoire de Paris, Section de Meudon, F-92195 Meudon Cedex, France}
\item{University of Durham, Department of Physics, South Road, Durham DH1 3LE,
U.K.}
\item{Laboratoire Leprince-Ringuet, IN2P3/CNRS, Ecole Polytechnique, F-91128 Palaiseau, France}
\item{APC, UMR 7164 du CNRS, 11 Place Marcelin Berthelot, F-75231 Paris Cedex 05, France}
\item{Dublin Institute for Advanced Studies, 5 Merrion Square, Dublin 2,
Ireland}
\item{Landessternwarte, K\"onigstuhl, D 69117 Heidelberg, Germany}
\item{Laboratoire de Physique Th\'eorique et Astroparticules, IN2P3/CNRS, Universit\'e Montpellier II, CC 70, Place Eug\`ene Bataillon, F-34095 Montpellier Cedex 5, France}
\item{Laboratoire d'Astrophysique de Grenoble, INSU/CNRS, Universit\'e Joseph Fourier, BP 53, F-38041 Grenoble Cedex 9, France}
\item{DAPNIA/DSM/CEA, CE Saclay, F-91191 Gif-sur-Yvette, France}
\item{Unit for Space Physics, North-West University, Potchefstroom 2520, South Africa}
\item{Laboratoire de Physique Nucl\'eaire et de Hautes Energies, IN2P3/CNRS, Universit\'es Paris VI \& VII, 4 Place Jussieu, F-75252 Paris Cedex 05, France}
\item{Institut f\"ur Theoretische Physik, Lehrstuhl IV: Weltraum und Astrophysik, Ruhr-Universit\"at Bochum, D 44780 Bochum, Germany}
\item{Institute of Particle and Nuclear Physics (IPNP), Charles University, V Holesovickach 2, 180 00 Prague 8, Czech Republic} 
\item{University of Namibia, Private Bag 13301, Windhoek, Namibia}
\item{European Associated Laboratory for Gamma-Ray Astronomy, jointly
supported by CNRS and MPG}
\end{enumerate}


\begin{sciabstract}
X-ray binaries are composed of a normal star in orbit around a neutron
star or stellar-mass black hole. Radio and X-ray observations have led
to the presumption that some X-ray binaries called microquasars behave
as scaled down active galactic nuclei. Microquasars have resolved
radio emission that is thought to arise from a relativistic outflow akin to
active galactic nuclei jets, in which particles can be accelerated to
large energies. Very high energy $\gamma$-rays produced by the
interactions of these particles have been observed from several active
galactic nuclei. Using the High Energy Stereoscopic System, we find evidence for gamma-ray emission $>$100 GeV from a candidate microquasar, \ls, showing that particles are also
accelerated to very high energies in these systems.
\end{sciabstract}


High resolution radio maps of X-ray binaries (XRB) have revealed
powerful outflows that are similar to the relativistic jets seen in
active galactic nuclei (AGN) \cite{Mirabel1994,Marscher2002}. In both
cases, the radio emission is due to synchrotron radiation from
particles accelerated to high energies.  These outflows probably
result from the accretion of material onto the compact object, albeit
on vastly different scales: the mass (size) of black holes in AGN is
at least $10^6$ times larger than that of compact objects in
XRB. Hence, XRB with resolved radio emission have been dubbed {\em
microquasars}, reflecting the suspicion of some fundamental scaling
with compact object mass.

The kinship should be most evident close to the black hole, where the
jet is launched and where the available energy reservoir to accelerate
particles is largest. In AGN the particles can reach energies such
that their non-thermal emission extends to the GeV-TeV $\gamma$-ray
regime, via Compton upscattering of ambient photons or as a result of
high energy hadron interactions. Because of relativistic bulk motion, this emission is most easily seen in blazars, where the AGN jet is
aligned close to the line-of-sight. Very high energy (VHE)
$\gamma$-rays are to be expected from some XRB if the physical
processes in the vicinity of the compact object are indeed
analogous. However, previous observations of VHE emission from XRB
were inconclusive \cite{Weekes1992}.

Two XRB have resolved radio emission in the 0.001 arcsecond range, which is presumed
associated with a relativistic jet, and possible counterparts in the
MeV-GeV domain \cite{Gregory1978,Taylor1992,Paredes2000}. LS~5039
(RX~J1826.2-1450) and LSI +61\deg303 (V615 Cas) are each composed of a
massive star in an eccentric orbit around an undetermined compact
object \cite{Hutchings1981,McSwain2004}. Their proposed $\gamma$-ray
counterparts are respectively localised to 0.5\deg\ (3EG~J1824-1514)
and 0.2\deg\ (3EG J0241+6103) in the Third Energetic Gamma-Ray
Experiment Telescope (EGRET) catalogue \cite{3EG}, so that the
association cannot be considered firm on the basis of positional
coincidence alone. The systems are quite inconspicuous in x-rays, with
low variability and luminosity of
$\sim\,10^{34}\,\mathrm{erg}\,\mathrm{s}^{-1}$ at 1 to 10~keV 
\cite{Bignami1981,Motch1997}, which is about one-tenth their
luminosity above 100~MeV, assuming that the EGRET sources are indeed
counterparts. The $\gamma$-ray spectra measured by EGRET are hard, with
photon indexes $\Gamma$ close to 2, which suggests that emission could extend
to the $>$100~GeV regime where atmospheric Cherenkov Telescope (ACT)
arrays operate. Constraining the emission cutoff energy provides
important clues as to the physics of the $\gamma$-ray
source. Furthermore, the XRB associations can be rigorously tested by
the superior angular resolution of ACTs.

Located in the Southern Hemisphere, \ls\ is ideally accessible to the
High Energy Stereoscopic System (\hess). \hess\ is an ACT array of
four telescopes located in Namibia, each equipped with a 107~m$^2$
mirror and a 960--photomultiplier tube camera
\cite{2003APh....20..111B,Hinton2004,2004APh....22..285F}. The
telescopes image the Cherenkov light from showers of particles created
when VHE $\gamma$-rays and cosmic rays enter the atmosphere. A central
trigger selects only showers seen by at least two telescopes. The
combination of high resolution imaging and stereoscopic shower
reconstruction allows efficient rejection of the background from
cosmic ray--initiated showers. The \hess\ 5 $\sigma$ sensitivity above
100~GeV reaches 1\% of the Crab Nebula flux after 25 hours of
observations close to the zenith. The direction of each $\gamma$-ray
shower is determined to 0.1\deg\ accuracy, which enables
localization of sources in the arc min range within the 5\deg\ field of view.

The Galactic Plane survey carried out by the full \hess\ array in the
summer of 2004 testifies to the performance of the instrument. In
\cite{Funk2005} we reported on the discovery of eight {\em extended}
VHE $\gamma$-ray sources within $\pm 30^{\rm o}$ of the Galactic
Center and $\pm 3^{\rm o}$ of the Plane. After standard quality
selection, a total of 25 pointings (10.5 h live time) taken during the
scan were found to cover the position of \ls. The data were
independently analysed with two separate calibration
pipelines\cite{2004APh....22..109A} and several different
reconstruction methods, all of which were in excellent agreement with
each other.  The results presented here are based on a maximum
likelihood adjustment of a shower model to the observed images to
obtain the direction, impact parameter, and energy of the
primary\cite{2003ICRC..2907}.  The likelihood adjustment also provides for each event a probability that is used to select the $\gamma$-ray--like events. An image size cut of 60 photoelectrons
was applied to avoid systematic effects due to inhomogeneities of the night sky background over the field of view, corresponding to an average post-cut spectroscopic energy
threshold of 220~GeV.

The reconstructed $\gamma$-ray map shows an excess 1\deg\ southwest of the
previously reported hotspot HESS~J1825-137 (Fig.~1). The excess on each position is calculated by comparing the number of source events, integrated over the instrument point spread function (PSF), against the estimated number of background
events in the same region.  We find a
significance of more than 7$\sigma$ for this new source, denoted HESS
J1826-148.  The source is point-like with a size upper limit of 50''
(1$\sigma$) given by a likelihood fit to a Gaussian source profile
folded through the detector response. This is actually the only point
like source discovered in the Galactic scan \cite{Funk2005}.  The best
position is $\alpha$(J2000)=18$^h$26$^m$15$^s$ and
$\delta$=-14\deg49'30'' (with statistical and systematic uncertainties of $\pm$32'' and 30'', respectively,  comparable to the uncertainties of the other survey
sources) \cite{Funk2005}. The positional accuracy is limited by the
presence of an extended nearby source and systematics in observations
taken at large offsets.  On-axis observations should improve the positional error to better than 15''.

The $\gamma$-ray spectrum was derived from the comparison of
reconstructed event energies (in a circle of 6 arc min around the source) to the prediction for a given spectral
shape \cite{Piron2001}.  The prediction uses energy resolutions and system acceptances
derived from simulations, taking into account the zenith angle
pointing of the array and the off-axis angle of the shower in the
field of view for each observation.  We find an acceptable fit
(chance probability of 7\% of getting a worse fit) to a power-law with a photon index
$\Gamma=2.12\pm0.15$ (Fig.~2). The low statistics currently limit
further investigation of more complex spectral shapes.  The average
integral flux above 250~GeV is $5.1 \times 10^{-12}~
\mathrm{photons}\,\mathrm{cm}^{-2}\,\mathrm{s}^{-1}$ (with statistical and systematic uncertainties of $\pm 0.8$ and $\pm1.3~\mathrm{photons}\,\mathrm{cm}^{-2}\,\mathrm{s}^{-1}$, respectively), corresponding to a
luminosity of $\sim 10^{33}$~erg~s$^{-1}$ at 3~kpc
\cite{Motch1997}. Errors on the spectral parameters correspond to
1$\sigma$ confidence interval. The highest energy measurement is at
$\sim 4\ \mathrm{TeV}$.

The positions of the supernova remnant G16.8-1.1 and the pulsar
PSR~B1822-14, which are both in the error box of the EGRET source and
are both plausible $\gamma$-ray sources, are inconsistent with the position of
HESS~J1826-148 (Fig.~1). Production of $\gamma$-rays from the
interaction of cosmic-rays with the interstellar medium is precluded
by the low H column density at the location of HESS~J1826-148 relative
to its surroundings \cite{Ribo2002}.  The radio position of \ls\ is
84'' away from the \hess\ position and well within the 3$\sigma$
confidence region (Fig.~1). We verified that there are no other radio or
X-ray sources compatible with HESS~J1826-148 in the NRAO VLA 1.4~GHz
Sky Survey \cite{NVSS} and in the XMM/Chandra fields analysed by
\cite{Martocchia2005}. The observations are not simultaneous with
those of H.E.S.S. so we cannot formally exclude a long episode of
flaring from a blazar. It would be surprising if this blazar were not detected in radio, as blazar emission at this wavelength is persistent at
levels of $\sim$ 100~mJy, well above the survey sensitivity (3 mJy for a point source). The present
evidence largely favors the association of \ls\ with HESS~J1826-148
and, by extrapolation, with the unidentified EGRET source 3EG
J1824-1514. The present H.E.S.S. data are consistent with a constant
flux (Fig.~S1). Confirmed $\gamma$-ray variability correlated with
other wavebands or a telltale modulation would fully establish the
association.

Several processes can lead to $\gamma$-ray emission in \ls. The bulk
of the luminosity in the system is emitted by the O6.5V stellar
companion ($L_\star$$\approx 10^{39}$~erg~s$^{-1}$) at an energy
$kT_\star \approx 3.5$~eV (Fig.~2). The binary separation varies from
$2R_\star$ to $6R_\star$ (where $R_\star$$\approx$7$\cdot 10^{11}$~cm), and
the radiation density reaches
$n_\star$$\approx$$10^{14}$~photons~cm$^{-3}$ close to the compact
object. These stellar photons can be boosted to $\gamma$-ray energies
by inverse Compton scattering on VHE electrons \cite{Bosch2004}. With
such radiation densities, the energy loss timescale for electrons in
the deep Klein-Nishina regime, giving a strict upper limit on the
radiative timescale, is $\sim $300~s. A short radiative timescale
relative to the escape timescale from the system ($\sim 100$~s)
implies that inverse Compton emission can be very efficient.

Accelerating electrons to the required energies may be hindered by
such rapid losses. Very high energies may be easier to reach for
protons, which suffer fewer radiation losses. VHE $\gamma$-rays may
then be emitted via proton-proton interactions with the stellar wind.  Assuming
a stellar wind with a mass loss rate of $10^{-6}$ M$_\odot$~yr$^{-1}$
and a velocity of $1000$~km~s$^{-1}$, the density at $2 R_\star$ is $\sim
10^{10}$ protons cm$^{-3}$. For such a density the $pp$ interaction
timescale is $\sim 10^5$~s. At least $10^2/10^5$=0.1\% of the protons
radiate for free streaming particles, implying a total kinetic energy of less than $10^{38}$~erg~s$^{-1}$. Protons may also interact with
stellar photons but the threshold is very high $\sim 10^{17}$~eV. The
$p\gamma$ timescale is $\sim 10^3$~s.

Photons emitted in the \hess\ energy range can interact with the
3.5~eV stellar radiation before leaving the system, producing $e^+e^-$
pairs. The cross-section maximum $\sigma_{\gamma\gamma}\approx
1.7\cdot 10^{-25}$~cm$^2$ occurs for $\gamma$-rays of energy $\approx
100$~GeV. The opacity is $\tau_{\gamma\gamma}= \sigma_{\gamma\gamma}
n_{\star} r \approx 20$ for a photon travelling a distance $r\approx
10^{12}$~cm (comparable to the binary separation). VHE photons emitted
close to the compact object are therefore always well inside the
`$\gamma$-photosphere' at which $\tau_{\gamma\gamma}\approx 1$. This
initiates an $e^+e^-$ pair cascade that redistributes the absorbed
radiation to lower frequencies. Gamma-rays at energies below 100~GeV
suffer little absorption because of the Wien cutoff of the stellar
spectrum. At higher energies the opacity decreases as 1/$E_\gamma$
\cite{Gould1967}. The VHE spectrum may therefore be hardened
relative to its intrinsic shape.

The absorption of TeV photons in the system can be mitigated. First,
the cross-section threshold and amplitude are angle-dependent, so that
$\gamma$-rays emitted in a cone pointing away from the companion are
not absorbed. Scattering of stellar photons in the wind will tend to
isotropize the radiation and diminish this effect. Variations are
expected because the geometry changes with orbital phase. Second,
$\gamma$-ray emission need not take place close to the compact
object. Observations of X-ray emission from XRB jets provide evidence
for acceleration of electrons to TeV energies on parsec scales
\cite{Corbel2002}. In \ls, acceleration at a shock $>$1 AU away from
the stellar companion would happen beyond the $\gamma$-photosphere.

The association of \ls\ with HESS~J1826-148 confirms that like some
AGN, XRB are able to accelerate particles to at least TeV
energies. Shocks from colliding ejecta or from jet - interstellar
medium interactions are natural candidates. Yet the association with
an outflow may be questioned in the absence of a direct detection of
relativistic motion in radio. The relativistic wind of a young pulsar
is a conceivable alternative for particle injection
\cite{Maraschi1981}. The situation would then resemble that in
PSR~B1259-63, a system composed of a radio pulsar in a much wider
3.4~yr eccentric orbit around a Be star. TeV emission from
PSR~B1259-63 was detected with \hess\ close to periastron \cite{PSRB}.
The higher wind density in \ls\ probably smears out any radio
pulses. The resolved radio emission from \ls\ would be due to
particles (cascade pairs) streaming out of the system. Further
insights into this system may be gained from combined radio and
$\gamma$-ray observations: Very Long Baseline Array maps attain a
spatial resolution of a few AU \cite{Paredes2000} tantalizingly close
to the $\gamma$-photosphere.


\bibliography{ls5039_f}

\begin{thebibliography}{10}

\bibitem{Mirabel1994}
I.~F. {Mirabel}, L.~F. {Rodriguez}, {\it Nature\/} {\bf 371}, 46 (1994).

\bibitem{Marscher2002}
A.~P. {Marscher}, {\it et~al.\/}, {\it Nature\/} {\bf 417}, 625 (2002).

\bibitem{Weekes1992}
T.~C. {Weekes}, {\it Space Science Reviews\/} {\bf 59}, 315 (1992).

\bibitem{Gregory1978}
P.~C. {Gregory}, A.~R. {Taylor}, {\it Nature\/} {\bf 272}, 704 (1978).

\bibitem{Taylor1992}
A.~R. {Taylor}, H.~T. {Kenny}, R.~E. {Spencer}, A.~{Tzioumis}, {\it Astrophys.
  J.\/} {\bf 395}, 268 (1992).

\bibitem{Paredes2000}
J.~M. {Paredes}, J.~{Mart{\'{\i}}}, M.~{Rib{\' o}}, M.~{Massi}, {\it Science\/}
  {\bf 288}, 2340 (2000).

\bibitem{Hutchings1981}
J.~B. {Hutchings}, D.~{Crampton}, {\it Publ. Astronom. Soc. Pacific\/} {\bf
  93}, 486 (1981).

\bibitem{McSwain2004}
M.~V. {McSwain}, {\it et~al.\/}, {\it Astrophys. J.\/} {\bf 600}, 927 (2004).

\bibitem{3EG}
R.~C. {Hartman}, {\it et~al.\/}, {\it Astrophys. J. Suppl.. Ser.\/} {\bf 123},
  79 (1999).

\bibitem{Bignami1981}
G.~F. {Bignami}, P.~A. {Caraveo}, R.~C. {Lamb}, T.~H. {Markert}, J.~A. {Paul},
  {\it Astrophys. J.\/} {\bf 247}, L85 (1981).

\bibitem{Motch1997}
C.~{Motch}, F.~{Haberl}, K.~{Dennerl}, M.~{Pakull}, E.~{Janot-Pacheco}, {\it
  Astron. Astrophys.\/} {\bf 323}, 853 (1997).

\bibitem{2003APh....20..111B}
K.~{Bernl{\" o}hr}, {\it et~al.\/}, {\it Astroparticle Physics\/} {\bf 20}, 111
  (2003).

\bibitem{Hinton2004}
J.~A. {Hinton}, {\it New Astronomy Review\/} {\bf 48}, 331 (2004).

\bibitem{2004APh....22..285F}
S.~{Funk}, {\it et~al.\/}, {\it Astroparticle Physics\/} {\bf 22}, 285 (2004).

\bibitem{Funk2005}
F.~{Aharonian}, {\it et~al.\/}, {\it Science\/} {\bf 307}, 1938 (2005).

\bibitem{2004APh....22..109A}
F.~{Aharonian}, {\it et~al.\/}, {\it Astroparticle Physics\/} {\bf 22}, 109
  (2004).

\bibitem{2003ICRC..2907}
M.~{de Naurois}, {\it et~al.\/}, {\it Proc. of the $28^{th}$ ICRC (Tsukuba)\/}
  (2003), p. 2907.

\bibitem{Piron2001}
F.~{Piron}, {\it et~al.\/}, {\it A\&A\/} {\bf 374}, 895 (2001).

\bibitem{Ribo2002}
M.~{Rib{\' o}}, {\it et~al.\/}, {\it Astron. Astrophys.\/} {\bf 384}, 954
  (2002).

\bibitem{NVSS}
J.~J. {Condon}, {\it et~al.\/}, {\it Astronomical Journal\/} {\bf 115}, 1693
  (1998).

\bibitem{Martocchia2005}
A.~{Martocchia}, C.~{Motch}, I.~{Negueruela}, {\it Astron. Astrophys.\/} {\bf
  430}, 245 (2005).

\bibitem{Bosch2004}
V.~{Bosch-Ramon}, J.~M. {Paredes}, {\it Astron. Astrophys.\/} {\bf 417}, 1075
  (2004).

\bibitem{Gould1967}
R.~J. {Gould}, G.~P. {Schr{\'e}der}, {\it Phys. Rev.\/} {\bf 155}, 1408 (1967).

\bibitem{Corbel2002}
S.~{Corbel}, {\it et~al.\/}, {\it Science\/} {\bf 298}, 196 (2002).

\bibitem{Maraschi1981}
L.~{Maraschi}, A.~{Treves}, {\it Mon. Not. R. Astron. Soc.\/} {\bf 194}, 1P
  (1981).

\bibitem{PSRB}
M.~{Beilicke}, M.~{Ouchrif}, G.~{Rowell}, S.~{Schlenker}, {\it IAUC\/} {\bf
  8300}, 2 (2004).

\bibitem{Tasker1994}
N.~J. {Tasker}, J.~J. {Condon}, A.~E. {Wright}, M.~R. {Griffith}, {\it
  Astronom. J.\/} {\bf 107}, 2115 (1994).

\bibitem{Ribo1999}
M.~{Rib{\' o}}, P.~{Reig}, J.~{Mart{\'{\i}}}, J.~M. {Paredes}, {\it Astron.
  Astrophys.\/} {\bf 347}, 518 (1999).

\bibitem{Marti1998}
J.~{Marti}, J.~M. {Paredes}, M.~{Ribo}, {\it Astron. Astrophys.\/} {\bf 338},
  L71 (1998).

\bibitem{Clark2001}
J.~S. {Clark}, {\it et~al.\/}, {\it Astron. Astrophys.\/} {\bf 376}, 476
  (2001).

\end{thebibliography}

\bibliographystyle{Science}


\begin{scilastnote}
\item The support of the Namibian authorities and of the University of Namibia
in facilitating the construction and operation of H.E.S.S. is gratefully
acknowledged, as is the support by the German Ministry for Education and
Research (BMBF), the Max Planck Society, the French Ministry for Research,
the CNRS-IN2P3 and the Astroparticle Interdisciplinary Programme of the
CNRS, the U.K. Particle Physics and Astronomy Research Council (PPARC),
the IPNP of Charles University, the South African Department of
Science and Technology and National Research Foundation, and by the
University of Namibia. We appreciate the excellent work of the technical
support staff in Berlin, Durham, Hamburg, Heidelberg, Palaiseau, Paris,
Saclay, and in Namibia in the construction and operation of the
equipment.
\end{scilastnote}

{\bf Supporting Online Material:} Fig. S1\\


\clearpage

\begin{figure}
\begin{center}
\includegraphics[width=\textwidth]{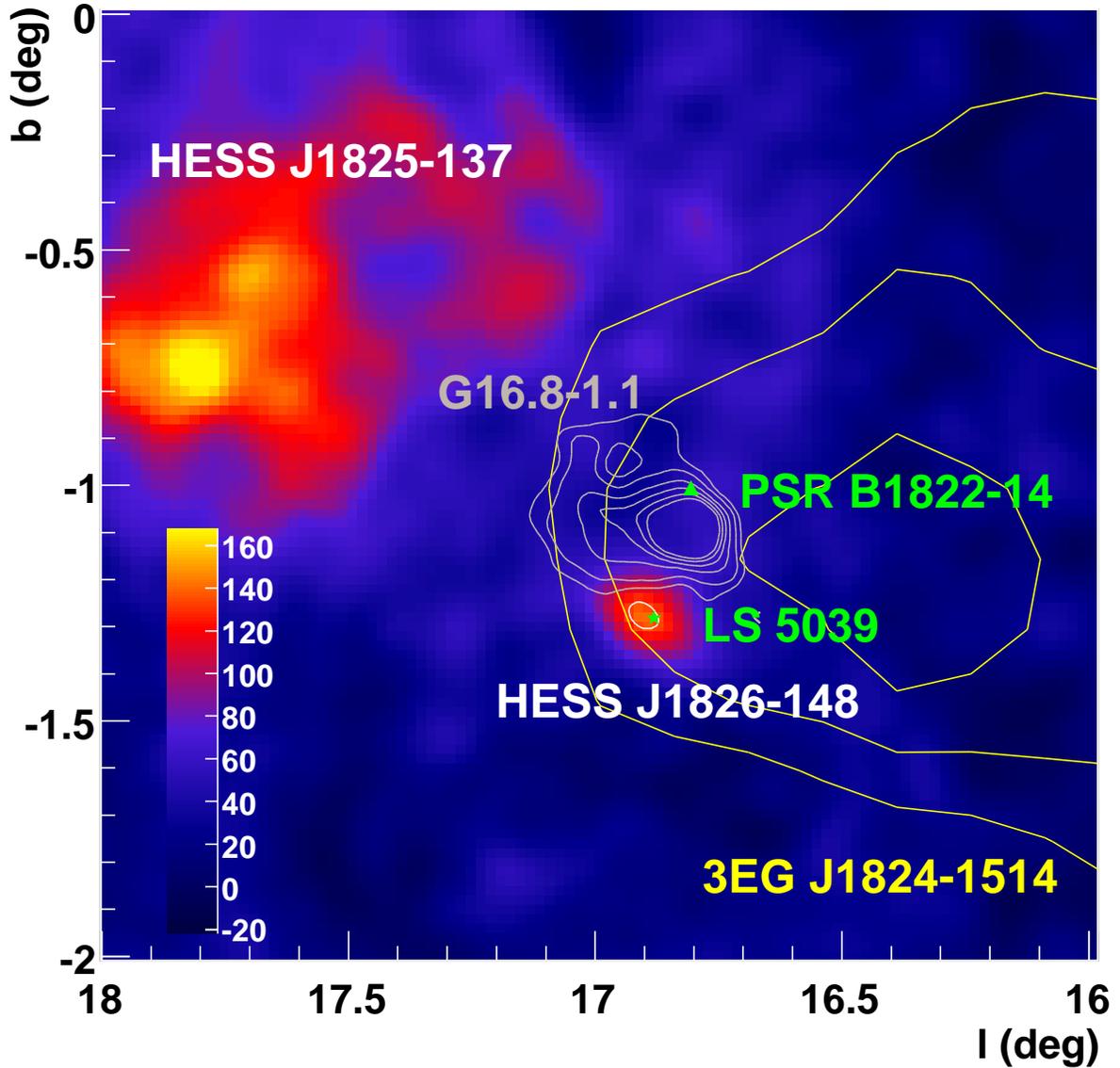}
\end{center}
\caption{Map of excess $\gamma$-ray emission in units of counts for the
region around \ls. The map has been smoothed by the point spread
function. The white ellipse shows the $3\sigma$ confidence region for
HESS~J1826-148. The radio emission from the SNR G16.8-1.1 is
represented by gray contours (0.05, 0.1, 0.2, 0.3, 0.4 and 0.5
Jy/beam) obtained from the Parkes-MIT-NRAO 6~cm radio survey
map\cite{Tasker1994,Ribo1999}.  The yellow contours show the 68\%,
95\% and 99\% confidence level region of the EGRET source
3EG~J1824-1514.  The green star marks the position of the radio source
associated with the microquasar LS~5039. HESS J1825-137 is discussed
in \cite{Funk2005}.}
\end{figure}

\begin{figure}
\begin{center}
\includegraphics[width=\textwidth]{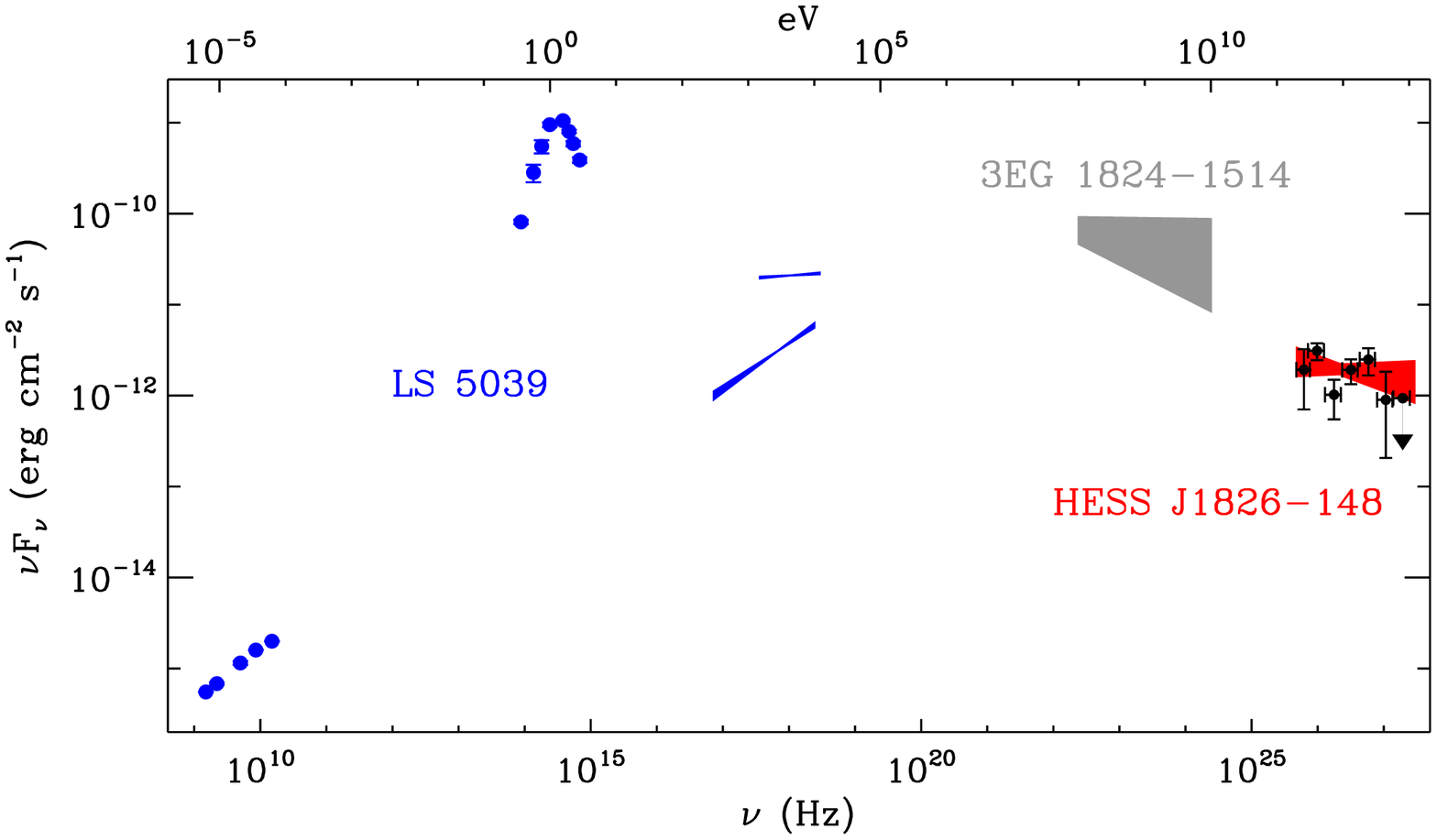}
\end{center}
\caption{Spectral Energy Distribution of LS~5039 including the
spectrum of HESS J1826-148 (points in black, power law fit in
red). The average radio, IR, optical and X-ray fluxes are shown in
blue \cite{Marti1998,Ribo1999,Clark2001,Martocchia2005}. Optical
fluxes are not dereddened. The two X-ray spectra correspond to the
historical 1998 high (Rossi X-ray Timing Explorer, RXTE) and 2003 low
(X-ray Multi-mirror Mission, XMM) flux observations of LS~5039. The
multi-year average flux above 100~MeV from the EGRET source
3EG~J1824-1514 is shown in gray \cite{3EG}.}
\end{figure}

\begin{figure}
\begin{center}
\includegraphics[width=\textwidth]{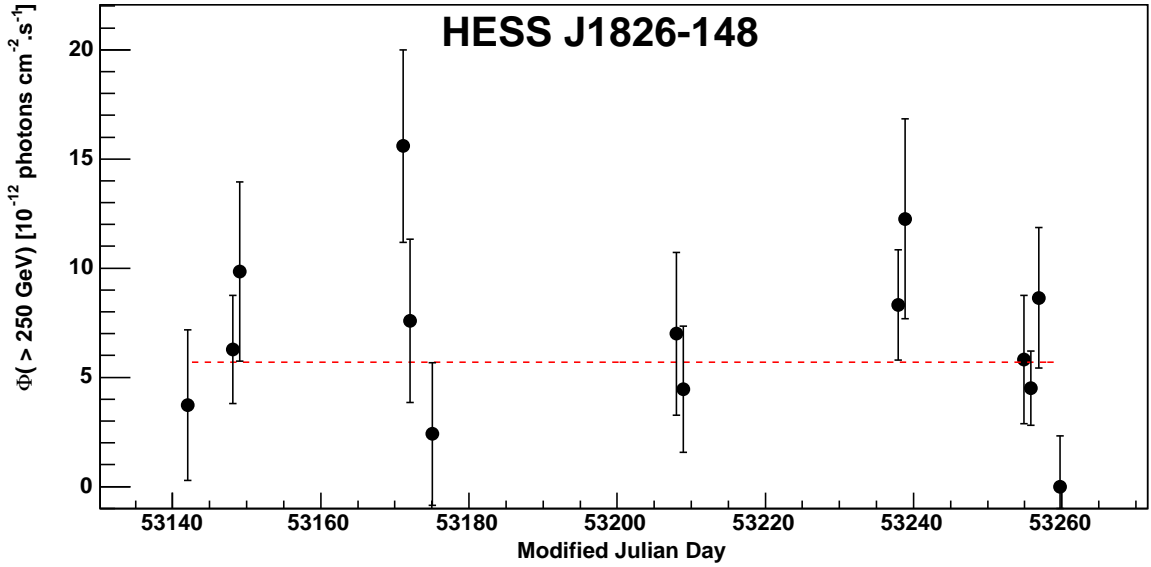}
\end{center}
\caption{ONLINE MATERIAL. Night by night integral flux above 250~GeV of HESS~J1826-148.
Taking into account a 20\% systematic uncertainty, the
hypothesis of a constant flux over the four months time span is
acceptable at the 15\% confidence level. The H.E.S.S. data covers
$\sim$ 27 cycles of the 4.4~day orbital period roughly uniformely in
phase. No periodic variations are apparent when folding the data using
the orbital ephemeris of \cite{McSwain2004}. The highest peak in a
2-100~day Lomb-Scargle periodogram of the run by run fluxes has a
significance $<$2~$\sigma$. The statistics are insufficient to study
daily variations in the spectral parameters.  }
\end{figure}

\end{document}